# Defining and Quantifying Conversation Quality in Spontaneous Interactions

Navin Raj Prabhu
Delft University of Technology
Delft, The Netherlands
lr.navin@yahoo.com

Chirag Raman
Delft University of Technology
Delft, The Netherlands
C.A.Raman@tudelft.nl

Hayley Hung
Delft University of Technology
Delft, The Netherlands
H.Hung@tudelft.nl

## ABSTRACT

Social interactions in general are multifaceted and there exists a wide set of factors and events that influence them. In this paper, we quantify social interactions with a holistic viewpoint on individual experiences, particularly focusing on non-task-directed spontaneous interactions. To achieve this, we design a novel perceived measure, the perceived Conversation Quality, which intends to quantify spontaneous interactions by accounting for several socio-dimensional aspects of individual experiences.

To further quantitatively study spontaneous interactions, we devise a questionnaire which measures the perceived Conversation Quality, at both the individual- and at the group- level. Using the questionnaire, we collected perceived annotations for conversation quality in a publicly available dataset using naive annotators. The results of the analysis performed on the distribution and the inter-annotator agreeability shows that naive annotators tend to agree less in cases of low conversation quality samples, especially while annotating for group-level conversation quality.

## 1 INTRODUCTION

Spontaneous interactions such as unplanned social conversations are typically non task-directed, unconstrained, and occur in natural situations [28, 30, 33]. In such interactions, the quality of the experience is a social construct that exists in the perception of individual participants. Such a subjective construct is generally quantified by relying on self-reported measures by the participants. However, such measures can suffer from biases from multiple sources—recall [24], social desirability [25], or egoistic[11, 25]. Furthermore, obtaining self reports might be precluded by privacy concerns. In this work we argue that the external perception of the quality of an interaction is an important construct towards the development of socially intelligent systems (e.g. social robots), as illustrated in Figure-1. In contrast to self-reported measures, a measure of perceived experience quantifies the individual or collective experiences of participants as perceived by a third-party observer [12, 21, 26]. Such measures are also resource efficient since they can be collected for existing datasets of spontaneous interactions where the participants are no longer available to provide self-reports. As such, externally perceived measures are more relevant to the development of artificial agents aimed at supporting and modulating human-human or human-robot interaction.

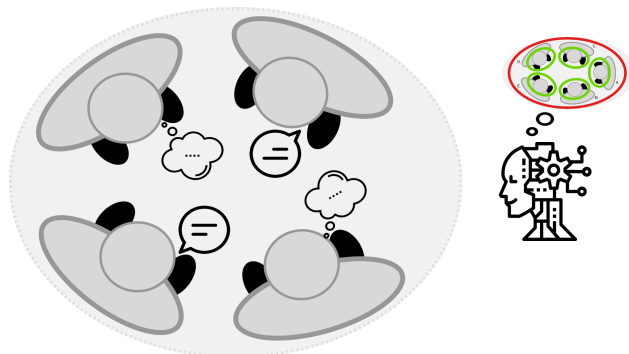

Figure 1: Illustration of individual experiences existing in the perception of interacting partners, and how a perceived measure of them are relevant for social robots.

Another challenge in the study of spontaneous interactions is that an experience of the social dynamics of an interaction is a multi-faceted construct. To quantify such an experience, it is very important to consider different overlapping aspects of the interaction: aspects such as interest-levels [12], involvement [26], cohesion [2], bonding [17], and rapport [23]. Existing literature in social psychology tends to consider such aspects in isolation. Consequently, an attempt to study the overall quality of individual experiences is still a knowledge gap.

In this work we make a two-fold contribution. Firstly, we introduce a novel measure of spontaneous interactions—*perceived conversation quality*. We formally define this construct and present its constituents by jointly considering overlapping aspects of the interaction. These aspects have thus far been considered only in isolation in social psychology literature. Secondly, we present an instrument in the form of questionnaires for collecting perceived annotations for *Conversation Quality*. We use the instrument to collect annotations of perceived conversation quality on a publicly available dataset of free-standing social conversations in-the-wild, and provide an analysis of the annotations. To the best of our knowledge, there is no existing work in the literature which has attempted to define and quantify the overall perceived quality of spontaneous interactions with respect to individual and group experiences.

The rest of the research paper is organized as follows. Firstly, in Section-2, we review several research works in existing literature to investigate the knowledge gap present and also draw inspirations to design the measure of perceived *Conversation Quality*. Secondly, in Section-3, we formally define the perceived *Conversation Quality* measure. Subsequently, in Section-4, we explain how Conversation Quality was quantified, with respect to its definition, using a publicly available dataset. In the same section, we also present and discuss the results of the analysis on the collected annotations. In Section-5, we discuss the several key findings and potential future works. Finally, in Section-6, we will conclude the research paper.

## 2 RELATED WORK

In this section, we present a literature review which discusses research works that have attempted to study social interactions. Firstly, we discuss how the study of social interaction are oprationalised by different researchers. Secondly, we discuss how different social constructs are quantified with their respective consideration and viewpoints. Finally, we present concluding remarks on existing literature and also discuss its existing knowledge gap.

### 2.1 Conversation Analysis

Fundamental research on social interactions was pioneered by Goffman [14], whose symbolic interaction perspective explains society via the everyday behavior of people and their interactions. Similarly, several other researchers have also operationalised the study of social interaction using different spatial and temporal aspects of the interactions.

Kendon (1990) [19, p.210], while studying the spatial aspect of face-to-face interactions, defined the f-formation system as, "the system of behavioural organisation by which certain spatial-orientational patterns are established and sustained in free-standing conversations". Similarly, Edelsky (1981), while studying the temporal aspect, examined a series of social interactions, and crucially observed two contrasting styles of conversation, the *exclusive* floors and the *cooperative* floors. According to Edelsky, the exclusive floor is characterised by a sense of orderliness, with only one person owning the floor at a time and turns rarely overlapping. In contrast, the cooperative floor is typified by a feeling of participants being "on the same wavelength" in a conversation that is a "free-for-all" ([10, p.384]), where there is a sense that no one owns the floor. The cooperative floor seems to capture the sense of cohesiveness and engagement that is associated with positive experiences in conversational scenarios.

Cooperative floors have been studied extensively in existing literature. For example, in the social sciences literature, measures of conversational equality and freedom [5, 21], measures of conversational fluency through frequent turns, turn overlap and turn duration [9], and measures of turn synchronisation [32], seem to resonate particularly strongly with Edelsky's views on cooperative floors. Spontaneous interactions are forms of such cooperative floors of interaction where there exists a sense of spontaneity amongst interacting partners and the interaction is non-task-directed. Reitter et al. (2010) [30] reveals the presence of contrasting behaviour patterns between a task-directed and a non-task-directed interaction. This motivates us to study such interactions separately with their respective considerations. In this research, we specifically concentrate on spontaneous non-task-directed interactions.

### 2.2 Quantifying Social Constructs

Spontaneous interactions are a dynamic social conversation setting, where a wide range of inter-personal relationships and social constructs emerge from within. Such relationships and constructs, as different aspects of individual experiences, have been studied extensively in existing literature. For example, social constructs which measure the inter-personal relationship (e.g. *Rapport* and *Bonding*), and social constructs which measure individual- and group- level behaviour (e.g. Involvement and Interest-levels) have been studied by researchers with their respective considerations.

Rapport and Bonding, as social construct, has been widely considered as a dyadic construct and as a self-reported measure [15, 17, 23]. Müller et al. (2018) [23] define Rapport as, "the close and harmonious relationship in which interaction partners are "in sync" with each other". The authors in their research, by considering Rapport as a dyadic-level phenomena, quantified Rapport between interacting pairs by relying on self-reported questionnaire measures adopted from Bernieri et at (1996) [3]. Another similar construct is Bonding, which measures the positive personal attachment, including "mutual trust, acceptance, and confidence" amongst interacting pairs [16]. Based on this definition, Jaques et al. (2016) [17], studied the Bonding in human-agent interactions. The authors in their research used the Bonding subscale of the Working Alliance Inventory (B-WAI) [16] to quantify Bonding in human-agent interactions.

While Rapport and Bonding tapped into the inter-personal relationships, several other social constructs, which quantify individual- and group- level behaviour, have also been studied in literature, e.g. Involvement, Engagement and Interest-levels. John H. Antil (1984) [1] defines involvement as "the level of perceived personal importance and/or interest evoked by a stimulus (or stimuli) within a specific situation". Following this definition, Oertel et al. (2011) [26] study participants' degree of involvement in social conversations. The authors in their research developed an perceived annotation scheme based on hearer independent, intuitive impressions and annotated for *ten* levels of involvement, each of the levels explaining its respective degree on involvement.

Similar to Involvement [26] and Engagement [27], as a perceived measure, Gatica-Perez et al. (2013) [12] define group interest-levels as, "the perceived degree of interest or involvement of the majority of the group". The authors relied on naive external annotators to annotate for interest-levels using audio-visual recordings of interactions, on a discrete 5-point scale. As instructions to annotators, the formal definition of group interest-level and examples of interest indicating activities (e.g. note-taking, focused gaze, and avid participation in discussion) were provided.

From the above discussed research works, we see that research works tend to quantify social constructs either by relying on self-reported measures or externally perceived annotation measures.

Self-report measures have many advantages, but they also suffer from specific disadvantages due to the way that subjects generally behave [11]. Self-reported answers may be exaggerated that respondents may be too embarrassed to reveal private details, various biases may affect the results, like social desirability bias [25]. In cases of series of short bursts of spontaneous interaction, subjects may tend to also forget longitudinal details and require an Experience Sampling Method (ESM) based data collection. At the same time, perceived measures are only an approximation of actual perceptions of the individuals and their perceptions. But, perceived measures are free from several keys issues faced by self-reported measures, mainly the issues of egoistic biases [11, 25], recall biases [24] and cognitive errors [24]. The characteristics of perceived measures, as discussed above, make them more suitable towards the development of social robots in dynamic spontaneous interactions.

On the other hand, from the above discussed research works, we see that there exists a common trend where researchers tend to focus on a particular aspect of social interactions and its individual experience, be it the inter-personal relationships or individual- and group- level engagement. Contrast to these studies, there have been some attempts to quantify individual experiences and social interactions with a holistic viewpoint by considering several unique aspects involved. Examples of such works include Cuperman and Ickes [8] and, Lindley and Monk [21]. However, these works considered different social settings, and were carried out for different purposes. While Cuperman and Ickes relied on self-reported measures in a dyadic clinical setting, Lindley and Monk quantified perceived enjoyment in a task-directed social interaction. Moreover, the set of aspects studied by the two works are mutually exclusive from one another and only capture a limited set of individual experiences in a spontaneous conversations.

Cuperman and Ickes (2009) [8], used the unstructured dyadic interaction paradigm to examine the effects of gender and the Big Five personality traits on the members' behaviors and perceptions of the interaction. For this purpose, the authors introduced the *Perception of Interaction* (POI) questionnaire to collected self-reported measures of a participant's perception of the interaction quality. This questionnaire contained 27 items that required the participants to rate their interaction experience, with respect to several unique aspects of the conversation. These aspects covered by POI include aspects such as, *Quality of the Interaction*, the *Degree of Rapport* they felt they had with the other person, and the *Degree to which they Liked* their interacting partner. This holistic measure of interactions has been successfully adopted by several other research to study social constructs such as bonding (Jaques et al. (2016) [17]) and interaction experience (Cerekovic et al. (2014) [6]) in human-agent interactions. Similar to Cuperman and Ickes [8], Lindley and Monk (2013) [21], with a holistic viewpoint on social interactions, studied several behavioral process measures to develop the *Thin Slice Enjoyment Scale* as a measure of experience and empathised enjoyment in social conversations. The thin-slice enjoyment scale specifically captures four unique aspects of a social interactions, namely *Conversation Equality*, *Conversation Freedom*, *Conversation Fluency* and *Conversation Enjoyment*.

## 3 DEFINING CONVERSATION QUALITY

In this section, we formally define the measure of perceived *Conversation Quality*. This measure, introduced in this research, has been inspired from Edelsky's definition of cooperative floors [10]. The cooperative floor, in contrast to exclusive floors, are self organising systems typified by a feeling of participants being "*on the same wavelength*" in a conversation that is a "*free-for-all*" [10, p.31]. This idea of the cooperative floor captures the sense of cohesiveness and engagement amongst interacting partners which is associated with positive individual experiences in social interactions. Considering spontaneous interactions as forms of cooperative floors, Edelsky's definition of cooperative floors [10, p.384] will be a suitable starting point to quantify the overall quality of spontaneous interactions.

With respect to Edelsky's definition of cooperative floors [10, p.384], in this research, we define the measure of *perceived Conversation Quality* in a spontaneous interaction as,

> the degree to which participants in the spontaneous interaction are of the **same wavelength** and maintain a **free-for-all** floor, as perceived by external observers.

The two keywords here, *same wavelength* and *free-for-all*, are the two high-level aspects of *Conversation Quality* and are vital in defining the measure. In a cooperative spontaneous interaction setting, the aspect of *same wavelength* is multi-faceted in nature and tends to capture the sense of cohesiveness, rapport and engagement that is associated with positive experiences in conversational scenarios. Similarly, the aspect of *free-for-all* intends to capture the equal opportunity shared amongst interacting partners in conversational scenarios. With these two high-level aspects of *Conversation Quality*, we believe the quantification of individual experiences in spontaneous interactions, with a holistic viewpoint, can be achieved. In the following sections, we will further discuss how these two high-level aspects of perceived *Conversation Quality* can be captured aptly using different constituents.

### 3.1 Constituents of Conversation Quality

From the literature review presented earlier (in Section-2), we see that previously studied social constructs, such as *cohesion* [2], *rapport*[23], *bonding*[17], *enjoyment*[22] and *interest-levels*[12] intend to capture a particular aspect of social interactions. Such social constructs do not intend to the quantify the overall quality of spontaneous interactions by capturing different aspects of individual experiences. For example, the measure of Rapport captures the *interpersonal relationship* in a social interaction by measuring the degree to which interacting partners are "*in-sync*" with each other" [23], but does not capture several other key aspects such as *degree of involvement* [26], *free-for-all* [21] and *interpersonal liking* [8]. Similarly, the measures such as *interest-levels* and *engagement* capture the *degree of involvement* displayed by individuals in the interaction, but does not capture the aspects such as *interpersonal relationship*[23], *quality of interaction*[8] and *free-for-all*[21]. In contrast to all these social constructs, the perceived Conversation Quality quantifies spontaneous interactions with holistic viewpoint.

In this section, we present the constituents of the measure of conversation quality. Each of these constituents intend to uniquely capture a specific aspect of individual experiences in a spontaneous interaction, thereby measuring the two high-level aspects of conversation quality, *same wavelength* and *free-for-all* aspects. Inspired by several research works in literature which study different aspect of individual experiences in spontaneous interactions, we present the *four* constituents of perceived Conversation Quality measure as, (1) Interpersonal Relationship, (2) Interpersonal Liking, (2) Nature of Interaction, and, (4) Equal Opportunity.

***Interpersonal Relationship***. The constituent of *Interpersonal Relationship* was designed by drawing inspirations from Jaques et al.'s Bonding [17] and Muller et al.'s Rapport [23]. This particular constituent of Conversation Quality captures the degree of association or acquaintance between interacting partners in a spontaneous interaction. The constituent directly measures interactions with respect to social constructs related to interpersonal relationships, e.g. rapport [23], cohesion [2] and bonding [17]. For example, the constituent measures the degree to which an individual was accepted and respected by other individuals in the group or the degree to which the other individuals were paying attention to the individual.

The *Interpersonal Relationship* amongst interacting partners is widely acknowledged to result in improved collaboration, and improved interpersonal outcomes, thereby having a key influence on the Conversation Quality and individual experiences.

***Interpersonal Liking***. The constituent of *Interpersonal Liking* was designed by drawing inspirations from Cuperman and Ickes' POI [8] and Cerekovic et al.'s Interaction Experience [6]. This particular constituent captures the degree to which an individual personally likes their interacting partners and the ongoing conversation with them. The constituent directly measures interactions with respect to the social constructs related to the interpersonal liking, like Cuperman and Ickes' Degree of Likeness [8] and Attraction [18]. For example, the extent to which an individual would like to interact more with their interaction partners or the extent to which an individual liked the other individuals in the interaction.

While the previously discussed constituent of Conversation Quality measured the interpersonal relationship based dimensional aspect of spontaneous interactions, this particular constituent measures another key aspect of such interactions, the Interpersonal Liking. While this particular constituent is key in quantifying spontaneous interactions, it is also important to note that this measure is an intimate measure of an individual's experience. Hence, similar constructs have been widely quantified by researchers using self-reported measures [8, 18]. With that in consideration, this particular constituent cannot be extended to perceived measures.

***Nature of Interaction***. The constituent of *Nature of Interaction* was designed by drawing inspirations from Cuperman and Ickes' POI [8]. This particular constituent of Conversation Quality directly captures the positive experiences and the nature of interactions amongst interacting partners. The constituent measures interactions with respect to the social constructs related to the positive experiences, like Cuperman and Ickes' Quality of Interaction [8] and Lindley and Monk's Empathised Enjoyment [21]. While the previously discussed constituents of Conversation Quality measured the interpersonal relationship and liking based dimensional aspects of spontaneous interactions, this particular constituent directly captures the nature of interaction amongst interacting partners and the positive experiences involved. For example, the degree to which the individual's interaction was smooth and relaxed or the degree to which the individual's interaction was forced and awkward.

***Equal Opportunity***. The constituent of *Equal Opportunity* was designed by drawing inspirations from Edelsky's work on cooperative floors [10]. This particular constituent directly captures the free-for-all concept in a spontaneous cooperative interaction, that is the equal opportunity shared amongst interacting partners. For example, free-for-all factors like conversation freedom [20], conversation equality [21] and an individual's opportunity to take the lead in the conversation [8, 17] resonate well with the concept of free-for-all and equal opportunity. Free-for-all is an essential aspect of cooperative floors and spontaneous conversations, and hence is an important constituent in measuring the *Conversation Quality*.

### 3.2 Perception of Conversation Quality

As discussed earlier, in this research we quantify conversation quality in spontaneous interactions using externally perceived measures. In this section, we present the forms in which *Conversation Quality* as a social construct can be perceived in spontaneous interactions.

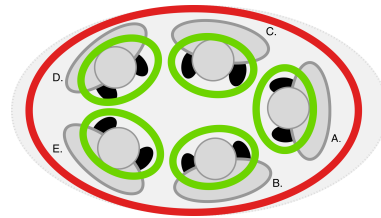

Figure 2: Illustration of the two forms of perceived *Conversation Quality*. The *red* and *green* boundaries illustrate the scope of observation to measure group-level and individual-level perceived Conversation Quality respectively.

Social interactions are multi-level systems that involve social constructs emerging from different levels of interactions [13]. For example, social constructs, in our case the *perception of Conversation Quality*, can emerge at different levels of interaction, e.g. individual level, dyadic level, group level or even the subgroup level. Perception of social constructs at different levels of interaction occur with a focus over the respective level. For example, an *individual-level* construct's perception occur with a prime focus on the individual and their interactions. When studying groups and teams, researchers can include individual-level and/or group-level phenomena in their research design. The ability of a socially intelligent system to perceive and understand both the individual-level and group-level *Conversation Quality* helps the system in understanding the influence of the individual-level phenomena on the group-level phenomena, which are key in development of several social robot applications.

In this research, we consider that the social construct of *Conversation Quality* exists in the perception of external observers in two forms - *Perceived Individual's experience of Conversation Quality* (the individual-level phenomena) and *Perceived Group's Conversation Quality* (the group-level phenomena). An illustration with respect to the two perceived measures of *Conversation Quality* and its scope of perception can be seen in Fig-2. The two forms of perceived conversation quality are defined in the coming sections.

***Perceived Group's Conversation Quality***. For this research, we define the perceived *group-level* conversation quality as an external observer's perception of the conversation quality of the group as a collection of all its individuals. This perceived measure directly taps into the what an external observer perceives or feels about the conversation going on in the whole group. On a high-level, this measure is the answer to the question,

> How do you rate *the overall quality of the conversation* involving the whole group, with respect to the group's *interpersonal relationship*, its *interpersonal liking*, its *nature of interaction* and the *equal opportunity* maintained in them.

The perceived annotation of this measure results in *one* rating per group, the group's externally perceived conversation quality.

***Perceived Individual's Experience of Conversation Quality***. For this research, we define the perceived *individual's* experience of conversation quality as an external observer's perception of the quality of the conversation as experienced by the *individual*. This perceived measure directly taps into an external observer's perception of an individual's experience in their conversation with the group. On a high-level, this measure is the answer to the question,

> How do you rate the *particular individual's experience* in the group, with respect to their *relationship*, their *liking*, their *nature of interaction* and the *equal opportunity* shared with their interacting partners.

The perceived annotation of this measure results in each individual in the group receiving a *Conversation Quality* rating given to them by the external observer. Hence, *n* perceived individual-level ratings are received per group, where *n* denotes the group-size.

### 3.3 Conversation Quality Questionnaires

In this section, we present the two questionnaires devised to measure the respective forms of *Conversation Quality* as perceived by external naive annotators. The. questionnaire can be used by naive external annotators to annotate for perceived *Conversation Quality* in non-task-directed spontaneous small group interactions, by relying solely on video clips of the interactions.

The two *Perceived Conversation Quality* questionnaires were devised by drawing inspirations from research works such as the *Perception of Interaction* (POI) by Cuperman and Ickes (2009) [8] and the *Thin-Slice Enjoyment Scale* (TES) [21]. The POI and TES questionnaires have been widely used by researchers to study social interactions in different scenarios [6, 17, 21]. Different from these studies, in this research we specifically focus on perceived social constructs in non-task-directed spontaneous group interactions. Hence, while drawing inspirations on the questionnaire items, we also modify the items to suit our social setting. The following steps were taken to modify the respective questionnaire items,

(1) All the items were made suitable for external annotators, suitable for perceived social constructs. That is, the items were modified to be directed towards the annotator themselves. For example, the item "I did not want to get along with the character" was modified to "The individual seemed to have gotten along with the group pretty well".

(2) All the items were modified to a small group social setting and not restricted to a dyadic interaction. For example, the question - "I felt accepted and respected by the character" was modified to - "The group members accepted and respected each other in the interaction".

(3) Questionnaire items which relied on the content of the conversation and ones which relied on modalities other than video clips were not considered. For example, the question which involved content of the conversation was excluded. e.g. the questionnaire item "The character often said things completely out of place" was excluded.

(4) Since our research focuses on a perceived measure of Conversation Quality, the intimate constituent of *Interpersonal Liking* was excluded while building the questionnaire. For example, the item "Did you desire to interact more with partner in the future?" was excluded as an external annotator cannot perceive an individual's personal liking.

The two questionnaires, devised to quantify perceived *Conversation Quality* at the individual- and group- level, can be found in the Appendix-7.1 and 7.2.

## 4 QUANTIFYING CONVERSATION QUALITY

In this section, we explain and discuss the strategy used to collect annotations for perceived conversation quality, using the *Perceived Conversation Quality* questionnaire presented earlier. The section is sub-sectioned as follows. Firstly, we will discuss the dataset used, secondly, we will discuss the procedure followed to collect perceived annotations for Conversation Quality, and finally, we will present the results of the analysis performed on the collection annotations.

### 4.1 Dataset

In this research, to quantify the perceived conversation in spontaneous interactions, we used the publicly available MatchNMingle dataset [4]. MatchNMingle is a multimodal dataset for the analysis of spontaneous free-standing conversational groups and speed-dates in-the-wild. The ecologically validated datasets contributes to the ecological validation of our study of conversation quality. The dataset leverages the use of wearable devices and overhead cameras to record a large number of in-the-wild social interactions during a real-life speed-date event and a cocktail party. For this research, we utilise only the data from the cocktail party.

The dataset consists of two hours of dynamic spontaneous interaction involving 92 participants, making it one of the largest dataset with a large number of participants and their ever-evolving interactions. This nature of the dataset was the prime motivation behind using the dataset for our study of spontaneous interactions. The interactions, in the dataset, were filmed using overhead GoPro cameras. In total 5 cameras were used to film the cocktail party (1080p, 30fps, ultra-wide field of view).

The MatchNMingle dataset also consists of f-formation annotations for 30-minutes cocktail party event. The annotations were performed using the visually perceivable spatial positions of participants during their interactions. The sub-sampled 30-minutes of annotated video segments were chosen randomly with an aim to eliminate the possible effects of acclimatization, and to maximize the density of participants and the number of social actions that could occur in the whole scene. In the 30-minutes segments of annotated f-formations in the mingling session, there we in total 174 f-formations. For this research, we consider a group to be an f-formation and the group members to be all the participants present in the particular f-formation. The duration of the f-formation interactions were distributed with a mean of 1.91, standard deviation of 2.13, median of 1.10 and a mode of 0.42. The distribution of the f-formation samples, with respect to its group size and duration, can be found in the Appendix, in Figure-3.

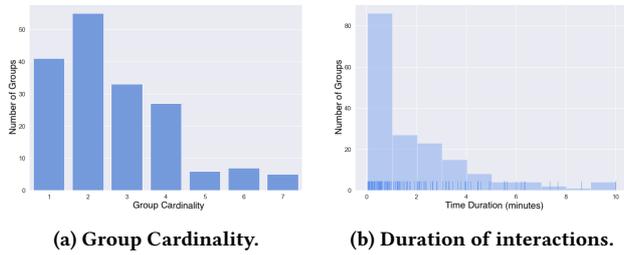

(a) Group Cardinality.  (b) Duration of interactions.

Figure 3: Distribution of f-formations in MatchNMingle.

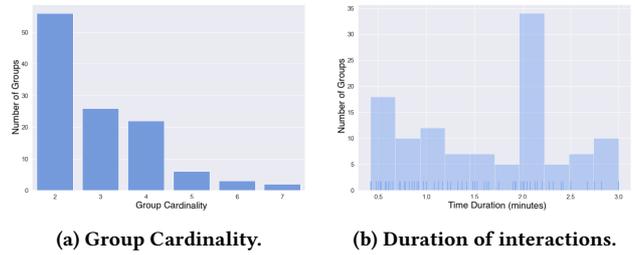

(a) Group Cardinality.  (b) Duration of interactions.

Figure 4: Distribution of final f-formation samples.

### 4.2 Annotation Procedure

In this subsection, we explain the annotation procedure used to collect annotations for perceived *Conversation Quality*. While explaining the annotation procedure, we also discuss several key considerations taken to devise the strategy.

The video clips of the spontaneous interactions, filmed using overhead cameras, was the only modality used for the manual annotation of the *Conversation Quality*. No audio data was used for the annotation process. Several other research works in literature have successfully collected rich annotation data by relying completely on video clips [12, 21]. Audio recordings in most of the conversation scenarios are unavailable due to privacy reasons. Moreover, annotations using audio data as one of the modality is also time consuming as they are generally prone to problems such as language constraints, audio noise, lack of clarity and sometimes requires speaker diarisation. On the other hand, manual annotations using only video recording are easier and less time consuming. At the same time, video recordings also have the capability to capture rich non-verbal behaviours of participants in the social interaction. Before using the f-formation groups for annotation, we cropped the respective f-formations out of the overhead video recordings. This was done in order to prevent annotators from getting distracted away from the current f-formation in focus.

Post cropping out f-formations from the video recordings, longer f-formation interactions were split into multiple smaller segments of interactions and then was presented to annotators as independent clips of social interactions. This was done in order to collect more reliable and granular annotations for longer group interactions. From Figure-3b, we see that the durations of f-formation interactions varies widely, from interactions of few seconds to that greater than 3-4 minutes. In that case, it is not reliable enough to have only one label annotation to define the conversation quality for the f-formation interactions of different durations. With the distribution in consideration, we decided to split f-formations of duration greater that 3 minutes into independent interactions of 1-2 minutes each. For the same reason for which we split the longer lasting f-formations, we omitted the f-formations with durations less than *30 seconds*. Post the omission and splitting processes, the total number of resulting f-formation groups was 115. The distribution of those groups with respect to the group size and interaction duration can be seen in Figures - 4a and 4b respectively.

With the processed video clips of spontaneous interactions, we decided to request naive annotators to help us in the annotation of *perceived Conversation Quality*. For this study, we were able to gather three naive annotators. The three annotators were aged between 22-30 years. Out of the three annotators, two were females and one was male. The annotators were provided with video clippings of the independent f-formations of spontaneous interactions and were asked to fill out both the *Perceived Conversation Quality* questionnaires (presented in Section-3.3). These f-formation groups were provided to the annotators in randomised order for each annotator, to prevent any annotator bias which might occur in case a strict f-formation clips order is followed.

### 4.3 Annotation Analysis

In this subsection, we present the results of the data analysis performed on the annotation responses collected through the annotation procedure explained earlier.

We first carried out principal component analysis on the annotations. This analysis showed that 71% and 65.2% of the variance, in the group-level and individual-level annotations respectively, could be explained by the first principal component. While the first four principal components are capable of explaining over 80% of the data variance. The eigenvalue bar-chart can be seen in Figure-5.

For further analysis of the annotations data, we plotted the data samples with respect to the first two principal components. The plot along with the factor loadings can be seen in Figure-6. Each line shown in the plots are the magnitude of loading of each question in the principal component space. A longer line indicates a larger variability of the vector in the two components and vice-versa. The numberings labelled on each loading line corresponds to the respective questionnaire item.

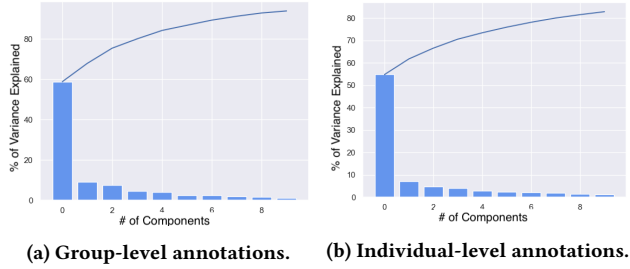

(a) Group-level annotations.　　(b) Individual-level annotations.

**Figure 5: Eigenvalue distribution (bar chart) and the cumulative percentage of the explained variability (line plot).**

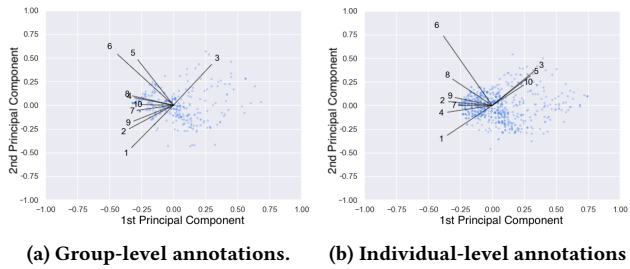

(a) Group-level annotations.　　(b) Individual-level annotations

**Figure 6: Plot of the factor loadings (black lines) and the samples (blue dots) in the first two principal components.**

From the annotations plots, at both the individual-level (Figure-6b) and group-level (Figure-6a), we see that questions are particularly clustered into two clusters, one cluster where questions show high variance towards the negative scale of the first principal component and the second cluster where questions show high variance towards the positive scale of the first principal component. On further analysis, we found that the questions in the two cluster corresponds respectively to the orientation of the scale for each question. For example, in the figure-6b, the questions 5, 3 and 10 are reversed in scale orientation from the rest of the questions. Similarly, in the figure-6a, the question 3 is reversed in scale orientation from the rest of the questions. This observation suggests that the three naive annotators treated the respective questionnaire items in a similar fashion. At the same time, we also see that few question items are strongly loaded with comparison to other items. For example, question 6 in both the annotations (6a, 6b) and also question 5 in (6a). It was interesting to note that, all these above mentioned highly loaded questions belong to the *Free-for-All* part of the questionnaire. This suggests us that the annotations for the free-for-all question items had the highest variance (between groups) in comparison with the other segments of the questionnaire.

### 4.4　Inter-annotator Agreement

Post the analysis on the annotation distributions, we performed analysis on the inter-annotator agreeability scores. For this, we used the *quadratic weighted kappa measure* [7], a variant of the Cohen's kappa measure. The measure allows disagreements to be weighted differently and is especially useful when the annotation data are ordinal in nature. To further analyse the final mean kappa agreements, for both the group- and individual- level annotations, we plotted the mean kappa score against the respective mean conversation quality score in a scatter plot, seen in Figure-7. A similar plot was used by Hung et al. [2] to analyse the inter-annotator agreeability for small-group meetings of different levels of cohesion.

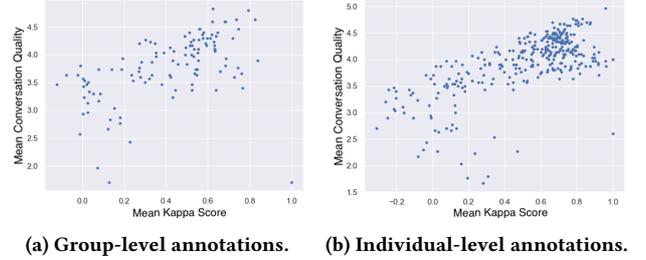

(a) Group-level annotations.　　(b) Individual-level annotations.

**Figure 7: Scatter plot between the Mean Kappa score ($\kappa$) and respective Mean Conversation Quality score.**

From the Figure-7, we see that there exists a linear relationship between mean kappa scores and mean conversation quality scores. That is, inter-annotator agreeability decreases as conversation quality scores decrease, suggesting that annotators agree better on conversations of higher quality when compared to conversations of lower quality. At the same time, a closer look reveals that, in the individual-level annotations (Figure-7b), there exists a small cluster of samples where annotators tended to agree higher for lower conversation quality samples as well. In contrary, for the group-level annotations (Figure-7a), annotators never agree well for low conversation quality samples. But, this was not expected by us. We expected similar results as seen in Hung et al.'s work [2], where inter-annotator agreements on cohesion levels for meetings were higher at the two extremes of the scale. Such a behaviour is seen only marginally and only for the individual-level annotations.

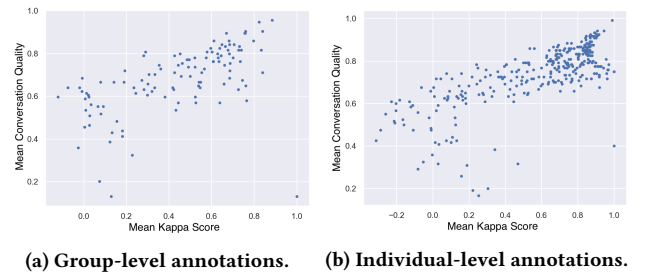

(a) Group-level annotations.　　(b) Individual-level annotations.

**Figure 8: Scatter plot between Mean Kappa score ($\kappa$) and Mean Conversation Quality score, after ZM adjustment.**

One widely used technique to handle low inter-annotator agreeability is to correct for mean-shifts, as used by Ringeval et al. (2013) [31]. The authors used a zero-mean (ZM) local normalization technique to remove an eventual bias in an annotator's annotations, e.g. a shift toward positive or negative values. We performed similar ZM adjustments on our annotations and similar analysis was performed. The resulting plots, seen in Figure-8, shows that no major changes are seen post the ZM technique. This suggests that there exists no mean shift between annotators but there exists a basic difference in annotator judgements.

## 5 DISCUSSION

Building on Edelsky's work on cooperative floors [10], we formally defined the measure of Conversation Quality and also presented its constituents. To further quantitatively study perceived Conversation Quality, we devised a questionnaire which measures, at the individual- and at the group- level, the perceived Conversation Quality in spontaneous interactions.

The questionnaires were used as an instrument to quantify Conversation Quality in a publicly available in-the-wild dataset of spontaneous interactions. By considering a stable f-formation to be a conversation group sample, we annotated the sample for perceived conversation quality at both the group- and individual- level. At the same time, works such as Raman et al. (2019) [29], prove existence of multiple conversation floors within an f-formation. With that in mind, as a future work, it would be interesting to quantify conversation quality for a particular conversation floor, rather than the whole f-formation.

Moreover, in this study, we have quantified a spontaneous interaction using one perceived Conversation Quality measure, assuming that there exists one stable perceived conversation quality score throughout the interaction. But, social interactions and individual experiences are dynamic in nature and requires a more fine grained approach. Several researchers have handled this using thin-slice based annotations [17][21]. Such a thin-slice based approach can help us further study the dynamics involved in the Conversation Quality of spontaneous interactions. Nevertheless, the devised questionnaire is flexible enough to be used for the collection of thin-slice annotations, and can be an interesting future work.

From the analysis on the distribution of the annotations, we saw that the naive annotators handled all the questionnaire items in a similar fashion. But, a deeper analysis with respect to the inter-annotator agreeability revealed that annotators tended to disagree with each other in cases of lower conversation quality samples. This behaviour was strongly prevalent in case of the group-level annotations than that of the individual-level annotations. A probable explanation, on the contrast between the two levels, could be that different naive annotators tend to employ different aggregation strategy to compile the overall group's conversation quality from individual-level measures, especially in cases of low conversation quality samples. With that in mind, as a future work, it would be interesting to use trained annotators in place of naive annotators. This could result in a richer dataset for further analysis and predictive modeling of perceived Conversation Quality.

## 6 CONCLUSION

In this paper, we designed a novel measure, the perceived *Conversation Quality*, which measures the overall quality of spontaneous interaction with a holistic view on individual experiences. To achieve this, we defined the measure to capture four unique aspects of social interactions, namely *Interpersonal Relationship*, *Interpersonal Liking*, *Nature of Interaction* and *Equal Opportunity*. Social interactions being multi-level systems, we defined that Conversation Quality can be perceived at two levels of perception, the individual-level (Perceived Individual's Experience of Conversation Quality) and group-level (Perceived Group's Conversation Quality).

To quantitatively study the novel measure, we devised two literature backed questionnaires which quantifies Conversation Quality at its respective levels of perception. We further used this questionnaire to collect perceived animations of Conversation Quality in a publicly available dataset, by relying on video clips of spontaneous interactions and naive external annotators. The analysis on the collected annotations revealed that, though the naive annotators treat the respective questionnaire items in similar fashion, they tend to agree less, with low inter-annotator agreement scores, in cases of low conversation quality samples. This behaviour is more prominent in group-level Conversation Quality annotations than that of the individual-level annotation, suggesting the usage of trained annotators in place of naive annotators.

Nevertheless, this research work is a pioneer in studying individual experiences in spontaneous interaction with a holistic viewpoint. Also, the devised questionnaire and the collected annotations can further facilitate the quantitative modeling of perceived Conversation Quality.

## ACKNOWLEDGMENTS

This research was partially funded by the Netherlands Organization for Scientific Research (NWO) under the MINGLE project number 639.022.606. We also thank Swathi Yogesh, Divya Suresh Babu, and Nakul Ramachandran for their time and patience in helping with annotating the dataset.

## REFERENCES

[1] John H Antil. 1984. Conceptualization and operationalization of involvement. *ACR North American Advances* (1984).
[2] Audio-visual Nonverbal Behavior, Hayley Hung, and Daniel Gatica-perez. 2010. Estimating Cohesion in Small Groups Using Audio-Visual Nonverbal Behaviour. *IEEE Transactions on Multimedia* 12, 6 (2010), 563–575. https://doi.org/10.1109/TMM.2010.2055233
[3] Frank J Bernieri, John S Gillis, Janet M Davis, and Jon E Grahe. 1996. Dyad rapport and the accuracy of its judgment across situations: A lens model analysis. *Journal of Personality and Social Psychology* 71, 1 (1996), 110.
[4] Laura Cabrera-Quiros, Andrew Demetriou, Ekin Gedik, Leander van der Meij, and Hayley Hung. 2018. The MatchNMingle dataset: a novel multi-sensor resource for the analysis of social interactions and group dynamics in-the-wild during free-standing conversations and speed dates. *IEEE Transactions on Affective Computing* (2018).
[5] Jean Carletta, Simon Garrod, and Heidi Fraser-Krauss. 1998. Placement of authority and communication patterns in workplace groups: The consequences for innovation. *Small Group Research* 29, 5 (1998), 531–559.
[6] Aleksandra Cerekovic, Oya Aran, and Daniel Gatica-Perez. 2014. How do you like your virtual agent?: Human-agent interaction experience through nonverbal features and personality traits. *Lecture Notes in Computer Science (including subseries Lecture Notes in Artificial Intelligence and Lecture Notes in Bioinformatics)* 8749 (2014), 1–15. https://doi.org/10.1007/978-3-319-11839-0_1
[7] J COHEN. 1968. Weighted kappa: nominal scale agreement with provision for scaled disagreement or partial credit. Psychology. *Bulletin, 70, 213â 220* (1968).
[8] Ronen Cuperman and William Ickes. 2009. Big Five Predictors of Behavior and Perceptions in Initial Dyadic Interactions: Personality Similarity Helps Extraverts and Introverts, but Hurts "Disagreeables". *Journal of Personality and Social Psychology* 97, 4 (2009), 667–684. https://doi.org/10.1037/a0015741
[9] Owen Daly-Jones, Andrew Monk, and Leon Watts. 1998. Some advantages of video conferencing over high-quality audio conferencing: fluency and awareness of attentional focus. *International Journal of Human-Computer Studies* 49, 1 (1998), 21–58.
[10] Carole Edelsky. 1981. Who's Got the Floor? *Language in Society* 10, 3 (1981), 383–421. http://www.jstor.org/stable/4167262
[11] John Garcia and Andrew R Gustavson. 1997. The science of self-report. *APS Observer* 10, 1 (1997).
[12] Daniel Gatica-Perez, Iain McCowan, Dong Zhang, and Samy Bengio. 2005. Detecting Group Interest-Level in Meetings. In *2005 IEEE International Conference*

on *Acoustics, Speech, and Signal Processing, ICASSP '05, Philadelphia, Pennsylvania, USA, March 18-23, 2005*. 489–492. https://doi.org/10.1109/ICASSP.2005.1415157
[13] Riemannian Geometry and Geometric Analysis. [n.d.]. *Advancing Multilevel Research Design: Capturing the Dynamics of Emergence - Steve*. Number Cdm. 581–615 pages.
[14] Erving Goffman. 1961. *Encounters: Two studies in the sociology of interaction*. Ravenio Books.
[15] Juan Lorenzo Hagad, Roberto Legaspi, Masayuki Numao, and Merlin Suarez. 2011. Predicting levels of rapport in dyadic interactions through automatic detection of posture and posture congruence. In *2011 IEEE Third International Conference on Privacy, Security, Risk and Trust and 2011 IEEE Third International Conference on Social Computing*. IEEE, 613–616.
[16] Adam O Horvath and Leslie S Greenberg. 1989. Development and validation of the Working Alliance Inventory. *Journal of counseling psychology* 36, 2 (1989), 223.
[17] Natasha Jaques, Daniel McDuff, Yoo Lim Kim, and Rosalind Picard. 2016. Understanding and predicting bonding in conversations using thin slices of facial expressions and body language. *Lecture Notes in Computer Science (including subseries Lecture Notes in Artificial Intelligence and Lecture Notes in Bioinformatics)* 10011 LNAI (2016), 64–74. https://doi.org/10.1007/978-3-319-47665-0_6
[18] Öykö Kapcak, Jose Vargas-Quiros, and Hayley Hung. 2019. Estimating Romantic, Social, and Sexual Attraction by Quantifying Bodily Coordination using Wearable Sensors. In *2019 8th International Conference on Affective Computing and Intelligent Interaction Workshops and Demos (ACIIW)*. IEEE, 154–160.
[19] Adam Kendon. 1990. *Conducting interaction: Patterns of behavior in focused encounters*. Vol. 7. CUP Archive.
[20] Catherine Lai and Gabriel Murray. 2018. Predicting group satisfaction in meeting discussions. *Proceedings of the Workshop on Modeling Cognitive Processes from Multimodal Data, MCPMD 2018* (2018). https://doi.org/10.1145/3279810.3279840
[21] Siân E. Lindley and Andrew F. Monk. 2013. Measuring social behaviour as an indicator of experience. *Behaviour & Information Technology* 32, 10 (oct 2013), 968–985. https://doi.org/10.1080/0144929X.2011.582148
[22] Florian Lingenfelser, Johannes Wagner, Elisabeth André, Gary McKeown, and Will Curran. 2014. An event driven fusion approach for enjoyment recognition in real-time. In *Proceedings of the 22nd ACM international conference on Multimedia*. 377–386.
[23] Philipp Müller, Michael Xuelin Huang, and Andreas Bulling. 2018. Detecting Low Rapport During Natural Interactions in Small Groups from Non-Verbal Behaviour. *CoRR* abs/1801.06055 (2018). arXiv:1801.06055 http://arxiv.org/abs/1801.06055
[24] Lance J. Rips Norman M. Bradburn and Steven K. Shevell. 1987. Answering Autobiographical Questions: The Impact of Memory and Inference on Surveys. In *New Series 1987*. 236(4798):157–167.
[25] David A Northrup. 1997. *The problem of the self-report in survey research*. Institute for Social Research, York University.
[26] Catharine Oertel, Céline De Looze, Stefan Scherer, Andreas Windmann, Petra Wagner, and Nick Campbell. 2011. Towards the Automatic Detection of Involvement in Conversation. In *Analysis of Verbal and Nonverbal Communication and Enactment. The Processing Issues*, Anna Esposito, Alessandro Vinciarelli, Klára Vicsi, Catherine Pelachaud, and Anton Nijholt (Eds.). Springer Berlin Heidelberg, Berlin, Heidelberg, 163–170.
[27] Catharine Oertel and Giampiero Salvi. 2013. A gaze-based method for relating group involvement to individual engagement in multimodal multiparty dialogue. In *Proceedings of the 15th ACM on International conference on multimodal interaction*. 99–106.
[28] Catharine Oertel, Stefan Scherer, and Nick Campbell. 2011. On the use of multimodal cues for the prediction of degrees of involvement in spontaneous conversation. *Proceedings of the Annual Conference of the International Speech Communication Association, INTERSPEECH* August (2011), 1541–1544.
[29] Chirag Raman and Hayley Hung. 2019. Towards automatic estimation of conversation floors within F-formations. In *2019 8th International Conference on Affective Computing and Intelligent Interaction Workshops and Demos (ACIIW)*. IEEE, 175–181.
[30] David Reitter, Johanna D Moore, and Frank Keller. 2010. Priming of syntactic rules in task-oriented dialogue and spontaneous conversation. (2010).
[31] Fabien Ringeval, Andreas Sonderegger, Juergen Sauer, and Denis Lalanne. 2013. Introducing the RECOLA multimodal corpus of remote collaborative and affective interactions. In *2013 10th IEEE international conference and workshops on automatic face and gesture recognition (FG)*. IEEE, 1–8.
[32] Leon Watts, Andrew Monk, and Owen Daly-Jones. 1996. Inter-personal awareness and synchronization: assessing the value of communication technologies. *International Journal of Human-Computer Studies* 44, 6 (1996), 849–873.
[33] D. Wyatt, T. Choudhury, and H. Kautz. 2007. Capturing Spontaneous Conversation and Social Dynamics: A Privacy-Sensitive Data Collection Effort. In *2007 IEEE International Conference on Acoustics, Speech and Signal Processing - ICASSP '07*, Vol. 4. IV–213–IV–216.

## 7 APPENDICES

The questionnaire items below have been organized in terms of the different constituents of Conversation Quality (Section-3.1). The numbering before each questionnaire item indicate the ordering of the items in the original questionnaire. The source for each term is provided at the end of each question.

### 7.1 Questionnaire for *Perceived Group's Conversation Quality*

<u>Instruction for the annotators:</u> Use the set of questions below to annotate your perception of the group's conversation quality, as seen in the video. Each interaction aspect in the below questionnaire should be rated using a five-point likert scale (Disagree strongly (1) to Agree strongly (5)). Read the questions carefully and observe the whole group carefully before annotating the video. You are allowed to re-watch the video again if required.

**Interpersonal Relationship**

4. The group members accepted and respected each other in the interaction. [8]
7. The group members seemed to have gotten along with each other pretty well. [8][17]
8. The group members were paying attention to their partners throughout the interaction. [8]
9. The group members attempted to get "in sync" with their partners. [8][17]
10. The group members used their partner's behavior as a guide for their own behavior. [8][17]

**Nature of Interaction**

1. The interaction within the group was smooth, natural and relaxed. [8]
2. The group members looked to have enjoyed the interaction. [8]
3. The interaction within the group was forced, awkward, and strained. [8]

**Equal Opportunity**

5. The group members received equal opportunity to participate freely in the interaction. [21]
6. The interaction involves equal participation from all group members. [21]

### 7.2 Questionnaire for *Perceived Individual's Experience of Conversation Quality*

<u>Instruction for the annotators:</u> Use the set of questions below to annotate your perception of the individual's experience in the conversation, as seen in the video. Each individual present in the conversation has to be annotated separately with the below questions. Each interaction aspect in the questionnaire below should be rated using a five-point likert scale (Disagree strongly (1) to Agree strongly (5)). Read the questions carefully and observe the individual carefully before annotating the video. You are allowed to re-watch the video again if required.

**Interpersonal Relationship**

8. The individual was paying attention to the interaction throughout. [8]

9. The individual seemed to have gotten along with the group pretty well. [8][17]

**Nature of Interaction**
1. The individual looked like they had a smooth, natural, and relaxed interaction. [8]
2. The individual looked like they enjoyed the interaction. [8]
3. The individual's interaction seemed to be forced, awkward, and strained. [8]
4. The individual looked like they had a pleasant and an interesting interaction. [8]
5. The individual looked uncomfortable during the interaction. [8]
10. The individual looked to be self-conscious during the interaction. [8]

**Equal Opportunity**
6. The individual attempted to take the lead in the conversation. [17][6]
7. The individual looked like they experienced a free-for-all interaction. [21]